\documentclass[a4paper,fleqn,usenatbib,mn2e,article,scrartcl]{mnras}

\usepackage{amsmath}
\usepackage{newtxtext,newtxmath}

\usepackage[T1]{fontenc}
\usepackage{ae,aecompl}
\usepackage{pdflscape}

\usepackage{graphicx}   
\usepackage{amsmath}    
\usepackage{amssymb}    

\usepackage{cases}
\usepackage{booktabs}

\title[Atom- Rydberg atom collisions: alkali]{The collisional atomic processes of Rydberg alkali atoms in Geo-cosmical plasmas}
\author[Ignjatovi{\'c} et al.] {
Lj. M. Ignjatovi{\'c}$^{1,2}$,
V. A. Sre{\'c}kovi{\'c},$^{1,2}$\thanks{E-mail: vlada@ipb.ac.rs}
and
M. S. Dimitrijevi{\'c}$^{2,3,4}$\\
$^{1}$University of Belgrade, Institute of Physics, P. O. Box 57, 11001 Belgrade, Serbia\\
$^{2}$Isaac Newton Institute of Chile, Yugoslavia Branch, Volgina 7, 11060 Belgrade, Serbia\\
$^{3}$Astronomical Observatory, Volgina 7, 11060 Belgrade 38, Serbia\\
$^{4}$Sorbonne Universit\'e, Observatoire de Paris, Universit\'e PSL, CNRS,LERMA,
F-92190 Meudon, France
}

\date{Accepted XXX. Received YYY; in original form ZZZ}

\pubyear{2017}

\begin{document}
\label{firstpage}
\pagerange{\pageref{firstpage}--\pageref{lastpage}}
\maketitle

\begin{abstract}
The chemi-ionization (CI) processes in atom-Rydberg-atom collisions are investigated in this contribution.
The rate coefficients for CI processes in $\textrm{Li}^{*}(n) + \textrm{Na}$,
$\textrm{Li}^{*}(n) + \textrm{Li}$, $\textrm{Na}^{*}(n) + \textrm{Na}$ and $\textrm{H}^{*}(n)+\textrm{Li}$
collisions are calculated for wide region of
temperatures $T \le 4000$ K and the principal quantum numbers $4 \le n\le 25$.
These processes are investigated, with a particular accent on the possibility of involving alkali metals
as the factors which may influence optical properties and modeling of weakly ionized layers of different stellar atmospheres.
The obtained rate coefficients are applied to models of the Io atmosphere.
Moreover, we present here further development of investigation of CI processes for other possible applications in spectroscopy such as in low temperature
laboratory plasma created in gas discharges, for example in microwave-induced
discharges at atmospheric pressure, where such plasma conditions may be favorable.
\end{abstract}



\section{Introduction}
\label{sec:intro}

Ionization processes which involve highly excited Rydberg atoms (RA) in different environments still attract attention of
scientists as they may be connected with the characteristics of many types of laboratory and astrophysical plasmas \citep[see, for example,][]{bez03, gne09, mih16}.
The study of RA with the development of laser technique has led to a great experimental advances
and in recent years renewed interest for such researches and enabled expansion in astrophysics.
The interpretation of line spectra with radiative transfer calculations requires spectroscopic data as well as collisional data (e.g., atomic
parameters, cross sections, rate coefficients, etc. \citealt{hau05,lin11}).
The fact that the ionization processes which involve highly excited atoms, like chemi-ionization (CI) processes \citep{mih12,oke12} influence the ionization level and atom excited-state populations, could influence the optical properties of the weakly ionized regions \citep{mih11} of alkali rich plasmas and potentially be important for the spectroscopy and modeling of such environment.

CI processes in collisions of excited alkali atoms with atoms in ground and
excited states were already considered, with a particular accent to the applications in stellar atmosphere modeling \citep{kly07} as well as for low temperature
laboratory plasma research created in gas discharges, for example in microwave-induced
discharges at atmospheric pressure, where plasma conditions \citep{yub07} may be favorable for processes investigated here.

There are several kinds of geo-cosmical plasmas where alkali atoms are present \citep[see e.g.][]{cha99,dup93}
and conditions for chemi-ionization in atom - Rydberg atom collisions are favorable, so that the corresponding data are useful for  modeling and
investigation of physical processes. We can mention cool stars, in particular brown and white dwarfs, lithium stars,
sodium clouds around Iovian satellite Io, cometary tails, and primordial gas containing Li atoms and ions.
As an example, we can cite \cite{deb12}, where is quoted the existence of metal-rich envelopes including Na around extremely
low-mass white dwarfs. Also, the mentioned chemi-recombination processes may be of interest for white dwarfs
polluted with metals by accretion from the surrounding environment (types DAZ, DBZ and DZ) and dusty white dwarfs. In fact, it is known
that in all dusty white dwarfs there is alkali atoms accretion onto their dusty disk. We note as well that white dwarfs
with the evidence of accretion can possess planetary systems \citep{gia14}, and the spectroscopic observations
of sodium atoms in these objects can contribute to the confirmation of this hypothesis. Namely, some of dusty  white dwarfs
have  powerful infrared excess produced by orbiting dust disk that contains planetary systems and planetesimals. Additionally,
these materials, containing alkali atoms as well, get accreted onto the white dwarf. Studying the accreted heavy elements,
including alkali atoms, becomes an effective way to measure directly the bulk compositions of extrasolar
planetesimals \citep{xu14}. Sodium atoms may be significant part of such heavy-element clouds so that detailed
investigation of Na emission features is of importance for investigating such astrophysical problems.
Recently, with combined Spitzer and ground-based Korea Microlensing Telescope Network observations
an Earth-mass planet orbiting ultra cool dwarf was identified and investigated \citep{shv17}. Consequently, data on Na atoms in such objects
can be very useful for determining planetesimals and Earth-mass planets in dusty disks surrounding
the central ultra cool dwarf. On the other hand, CI
processes modify plasma characteristics which influences spectral properties as well, so that their investigations contributes to
better modelling and analysis of such plasmas.

Study of CI processes may be of interest for the investigation and modelling of plasma in cold lithium stars,
intensively discussed in the literature \citep[see][]{nor98, sha01, sha03}. As an example, recently
\cite{li18} found high-lithium abundance in newly discovered 12 low-mass, metal-poor, main-sequence stars, and in
red giant stars in the Milky Way halo. Collisional processes including CI could be of interest for
studying of lithium-rich stellar atmospheres as an additional canal for the creation of Li I atoms. The considered
CI processes influence as well the ionization level and atom excited-state populations,
so that they could affect the optical properties of the weakly ionized regions of alkali rich plasmas.
Additionally, already in \cite{bro74a} the first neutral sodium cloud near Jovian
satellite Io has been detected \citep[see also paper][]{bro74b,feg00}. Since the investigation of these sodium clouds are necessary to better understand
the interaction between Io's atmosphere and Jovian magnetosphere and the processes in the Jovian
surroundings \citep{men90,wil02}, the data on CI processes during alkali RA collisions may be of interest.
In a recent paper \cite{sch05}, authors estimated and expected abundance of lithium at Io.

Another type of plasma interesting for the application of present results is in cometary tails, where there is  possible significant
presence of alkali atoms, especially of sodium atoms \citep{cre02}. The existence of the neutral tail containing sodium
was confirmed through observations of comet Hale Bopp in \cite{cre97}.

It is also known that in the early Universe the chemical composition of the primordial gas contained lithium,
and during the cooling of Universe there is an epoch when conditions for presently investigated RA collisional
processes were favorable \citep{puy07}. Collisional recombination, ionization and Rydberg states can play important role for the
early Universe chemistry \citep{cop11}.
Consequently, data on rate coefficients for CI processes in atom - Rydberg atom collisions of Na and Li atoms,
may be of interest for a number of astrophysical plasmas, as well as for the early Universe chemistry.

Finally, the properties of alkali plasmas are of basic interest and of importance for some laboratory and technical applications \citep{hor15,pic17}.
For instance, investigation of the fluorescence spectrum of sodium atoms is important in spectroscopy and laboratory investigation \citep{efi17}.
Also, lithium is an alkali metal of considerable technological interest. It is used in new light sources, solar thermal power plants, inertial confinement fusion reactor, electrochemical energy sources, next-generation microelectronics, etc \citep{stw80, gro17}.
\begin{figure}
\begin{center}
\includegraphics[width=\columnwidth,
height=0.68\columnwidth]{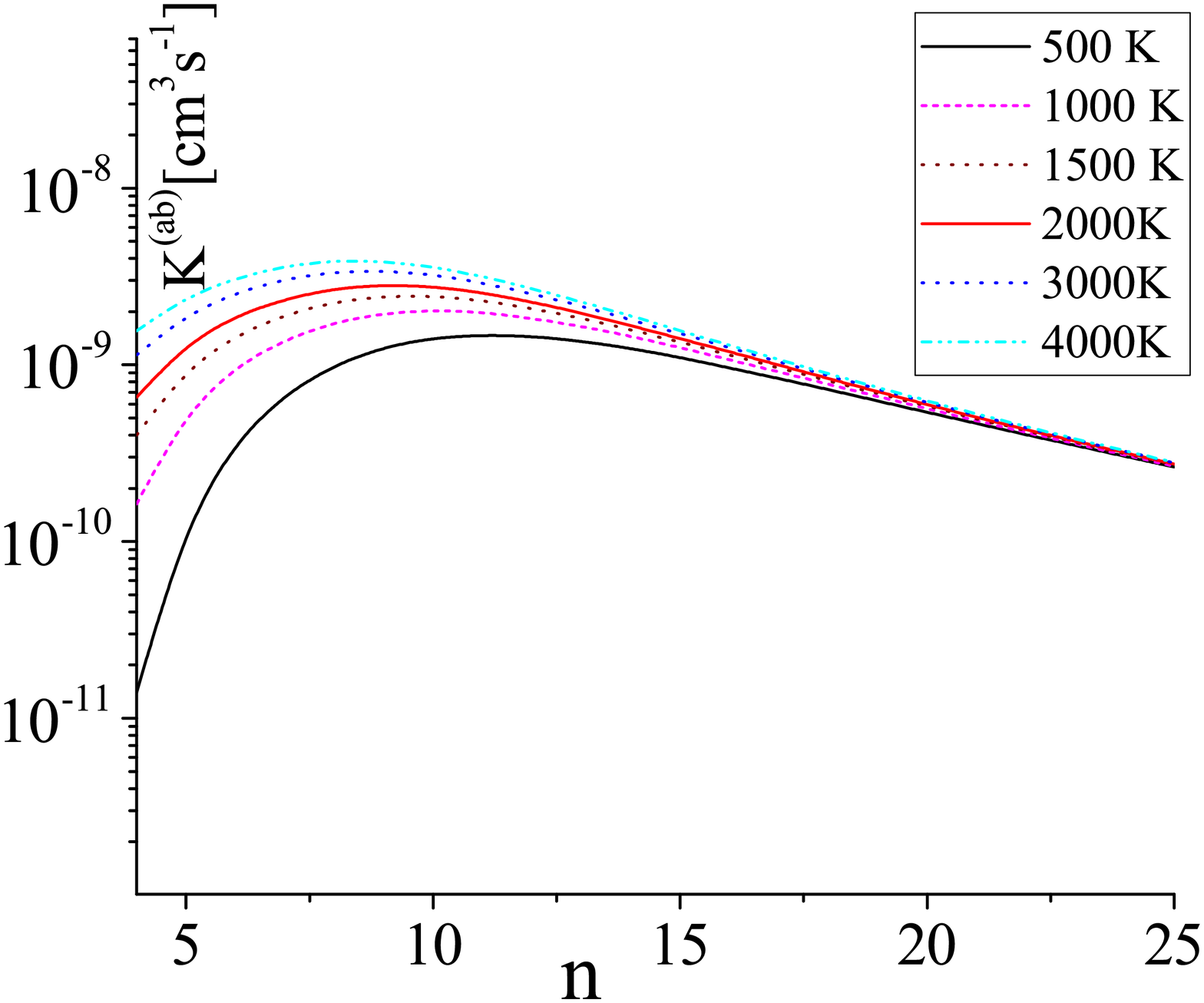}
\includegraphics[width=\columnwidth,
height=0.68\columnwidth]{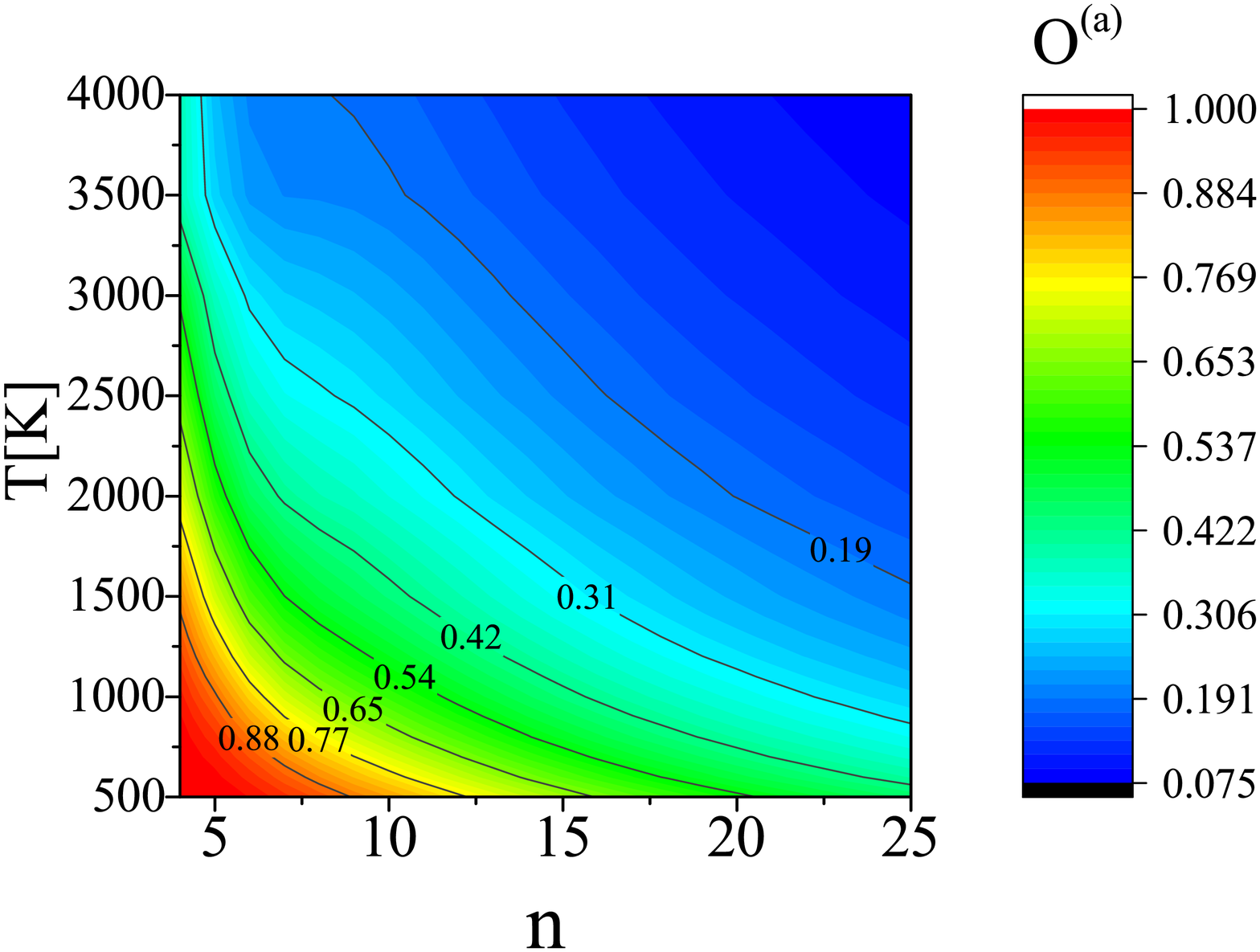}
\caption{Up: The total rate coefficient $K^{(ab)}(n,T)$, Eq. (\ref{eq:Kn1})
for CI processes (\ref{eq:nsim1}) and (\ref{eq:nsim2}) in $\textrm{Li}^{*}(n) + \textrm{Li}$ collisions.; Down: The coefficient $O^{(a)}(n,T)$, Eq. (\ref{eq:X})
for CI in $\textrm{Li}^{*}(n) + \textrm{Li}$ collisions.}
\label{Fig:2}
\end{center}
\end{figure}

\section{Theoretical remarks}

In this contribution we studied two types of CI processes: the non-symmetric processes
\begin{subequations}\label{eq:nsim}
\begin{numcases}{A^{*}(n) + X  \Rightarrow e +}
   AX^{+}, \label{eq:nsim1}
   \\
   A + X^{+}, \label{eq:nsim2}
\end{numcases}
\end{subequations}
and the symmetric processes
\begin{subequations}\label{eq:sim}
\begin{numcases}{A^{*}(n) + A  \Rightarrow e +}
   A_{2}^{+}, \label{eq:sim1}
   \\
   A + A^{+}, \label{eq:sim2}
\end{numcases}
\end{subequations}
where $A$, $X$, $A^{+}$ and $X^{+}$ are atoms and their atomic ions in the
ground states, $A^{*}(n)$ is the atom in a highly excited (Rydberg)
state with the principal quantum number $n\gg1$, i.e. RA, $A_{2}^{+}$ and
$AX^{+}$ are the molecular ions in the ground electronic states.
In Eq. (\ref{eq:nsim}) the ionization
potential $I_X$ of the $X$ atom is less than the ionization potential
$I_A$ of the atom $A$. The CI reactions (\ref{eq:nsim}) and (\ref{eq:sim}) can be divided
according to the products of the reaction: to the channel (a) of associative ionization i.e. creation of molecular ions $A_{2}^{+}$ or $AX^{+}$
and to the non-associative ionization channel (b).

In \cite{ign05} and \cite{ign08}, the processes $A^{*}(n) + A$ and $A^{*}(n) + X$,
with Li and Na have been considered for laboratory plasma for the conditions of single-beam and crossed-beam
cases for the temperature range $600 \textrm{ K} \le T \le 1100 \textrm{ K}$.
Here we further investigate CI processes for astrophysical plasma and give results which enable modeling of geo-cosmic
weakly ionized plasma for wide range of plasma parameters ($500 \textrm{ K} \le T \le 4000$ K and $4 \le n\le 25$).
and for the first time examine the CI processes in $\textrm{H}^{*}(n)+\textrm{Li}$ collisions.

\textbf{The calculated quantities:}
The method of the dipole resonant mechanism is used for the calculations of
the rate coefficients of the processes (\ref{eq:nsim}) and (\ref{eq:sim}). We present here the brief
description with the basic theory \citep[for details see e.g. papers][]{mih12,oke12}.
The calculations of these rate coefficients are performed for the principal quantum
number $4 \le n \le 25$ and temperatures up to $ 4000 \textrm{ K}$.
The results of calculation, as well as the necessary discussion, are
presented in Sec. \ref{sec:results}

The cross section for associative ionization
$\sigma^{(a)}(n,E)$ (channels \ref{eq:nsim1} and \ref{eq:sim1}) and the total CI
cross-section $\sigma^{(ab)}(n,E)$ (for \ref{eq:nsim} and \ref{eq:sim})
can be presented in the form
\begin{equation}
\sigma^{(a,ab)}(n,E) = 2 \pi
\int\limits_{0}^{\rho_{m}^{(a,ab)}(E)} P^{(a,ab)}(n,\rho,E)
\rho d\rho.
\label{eq:sigma}
\end{equation}
Here $P^{(a,ab)}(n,\rho,E)$ is the CI probability which
describes the processes separately (a)and together (ab), $\rho$ is the parameter of impact,
$\rho_{m}^{(a, ab)}(E)$ are the upper limits of corresponding impact parameters
\citep[for details see][]{ign05,ign08}.
The partial rate coefficients $K^{(a)}(n,T)$ and $K^{(b)}(n,T)$ and the total ones $K^{(ab)}(n,T)$ can be presented with
\begin{equation}
K^{(ab)}(n,T)=K^{(a)}(n,T)+K^{(b)}(n,T).
\label{eq:Kzbir}
\end{equation}
By definition, rate coefficients $K^{(ab)}(n,T)$ and
$K^{(a)}(n,T)$ are given by relations
\begin{equation}
\begin{split}
K^{(a)}(n,T) =
\int\limits_{0}^{E_{m}^{(a)}(n)} v  \cdot \sigma^{(a)}(n,E)
f(v;T) dv, \\ K^{(ab)}(n,T) = \int\limits_{0}^{\infty} v \cdot
\sigma^{(ab)}(n,E) f(v;T) dv, \label{eq:Kn1}
\end{split}
\end{equation}
where the cross-sections $\sigma^{(a,ab)}(n,E)$ are determined
by Eq.~(\ref{eq:sigma}), $E_{m}^{(a)}(n)$ is the
upper limit of $E$ relevant for the associative
ionization process (\ref{eq:nsim1}) or (\ref{eq:sim1})  (see \cite{ign05,mih12}), and $f(v;T)=f_{cell}(v;T)$ is Maxwell
distribution function $f(v;T) = 4 \pi \left[\frac{M}{2 \pi kT}\right]^{3/2} v^{2} e^{-Mv^{2}/2kT},$
where $M$ is reduced mass of sub-system $A^{+}+A$ or $A^{+}+X$.
The rate coefficient
$K^{(b)}(n,T)$ for the process (b) is determined
from Eq.~(\ref{eq:Kzbir}).
The potential curves of the molecular ions, the
square of dipole matrix element for the transition between states,
as well as other needed parameters can be found  in \cite{ign05}, \cite{ign08} and \cite{ign14}.

The relative contribution of the associative processes (\ref{eq:nsim1}), (\ref{eq:sim1}) and
non-associative processes (\ref{eq:nsim2}), (\ref{eq:sim2}) can be characterized by the corresponding
coefficients
\begin{equation}
O^{(a)}(n,T)=\frac{K^{(a)}(n,T)}{K^{(ab)}(n,T)},
\quad O^{(b)}(n,T)=\frac{K^{(b)}(n,T)}{K^{(ab)}(n,T)} .
\label{eq:X}
\end{equation}

The results of the rate coefficients of all investigated CI processes in this work are given in tabulated form in the online version of this article.
\begin{figure}
\begin{center}
\includegraphics[width=\columnwidth,
height=0.68\columnwidth]{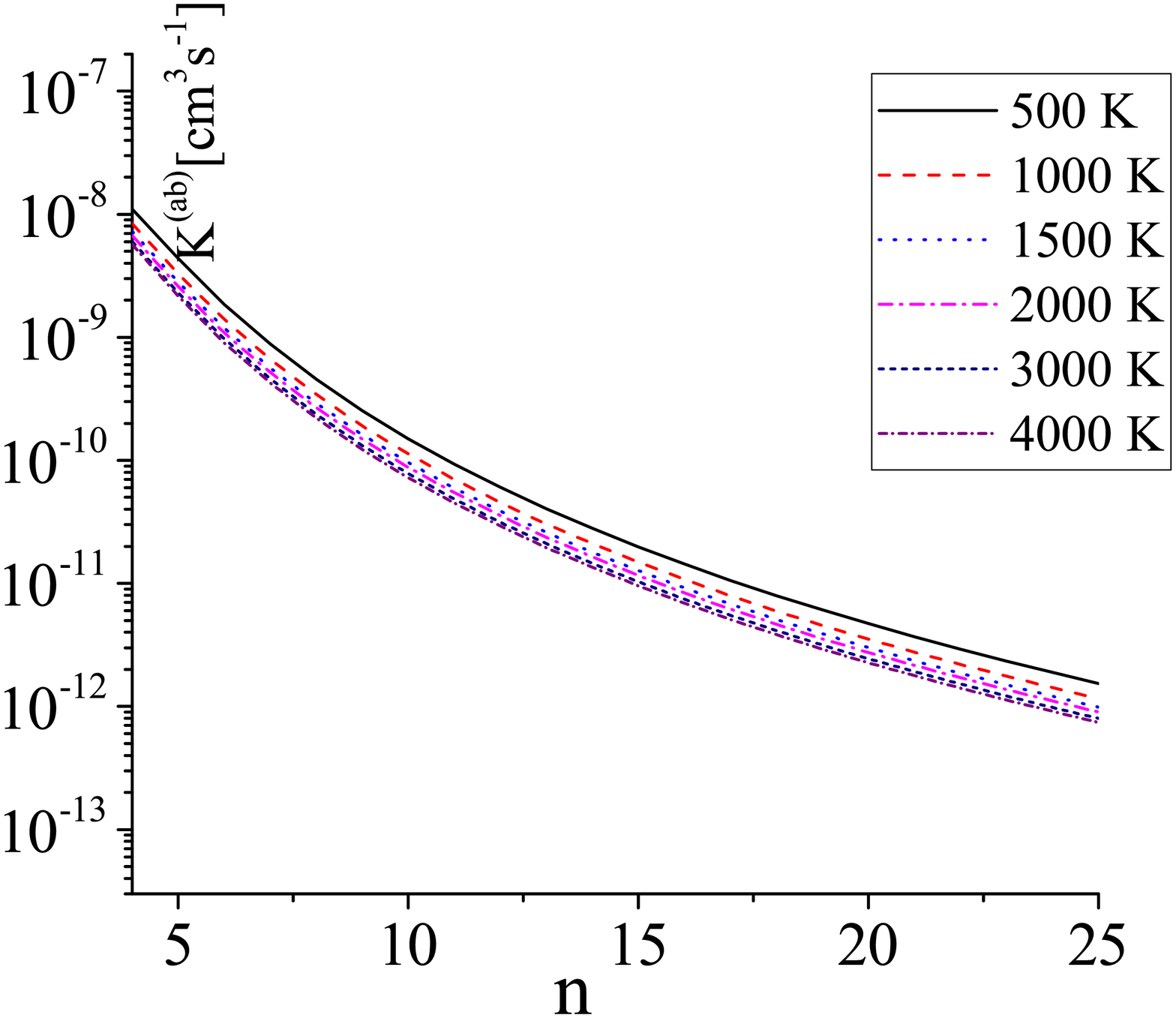}
\includegraphics[width=\columnwidth,
height=0.68\columnwidth]{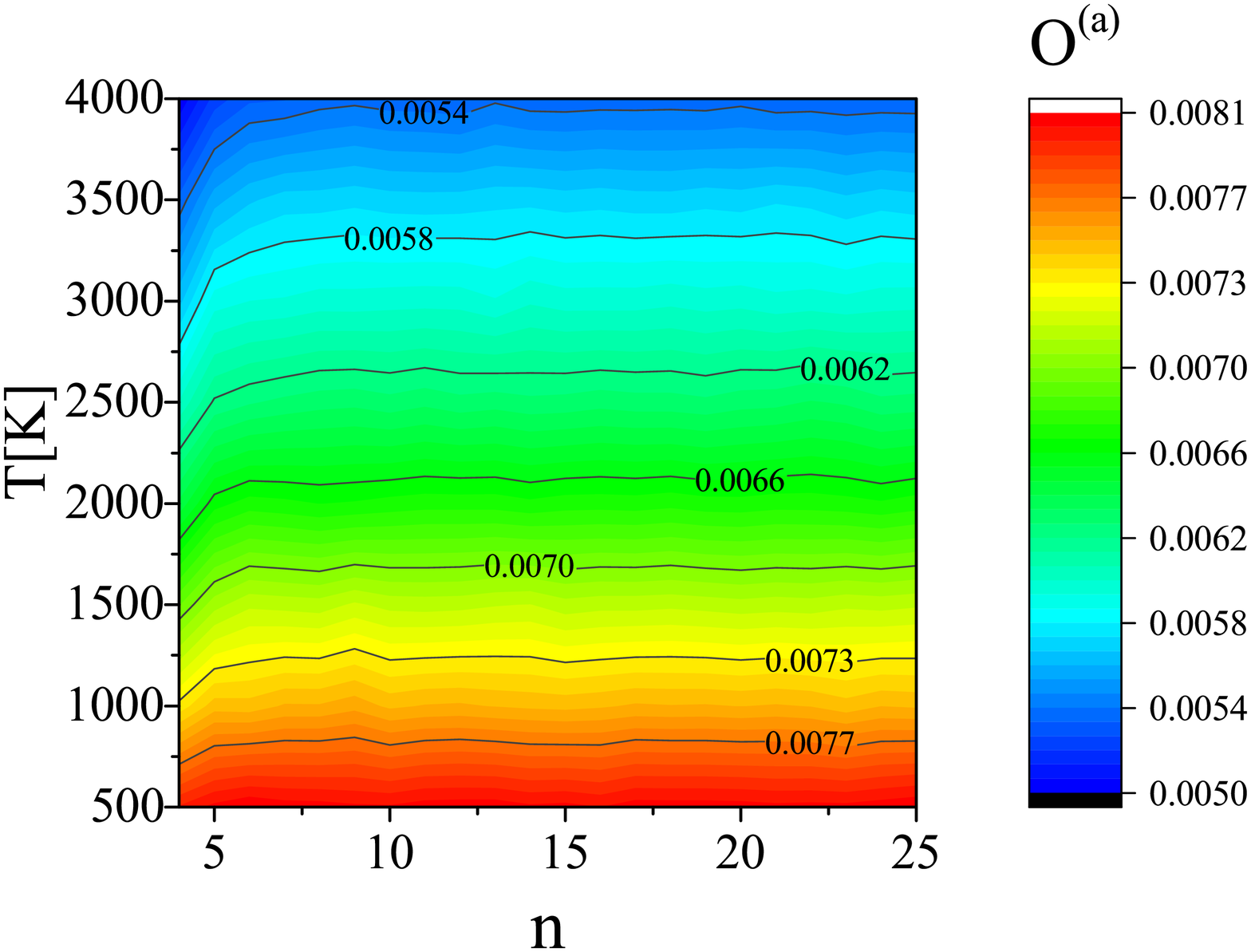}
\caption{Same as in Fig \ref{Fig:3} but for $\textrm{H}^{*}(n) + \textrm{Li}$ collisions.}
\label{Fig:3}
\end{center}
\end{figure}

\begin{table*}
\centering
\caption{The fits of the  Eq.(\ref{eq:Fit}) to the rate coefficient. A portion is shown here for guidance regarding its form and content.}
	\label{tab:fit}
\begin{tabular}{ccccccccccccc}
\hline
   & \multicolumn{3}{c}{$\textrm{Li}^{*}(n) + \textrm{Li}$}        & \multicolumn{3}{c}{$\textrm{Li}^{*}(n) + \textrm{Na}$}        & \multicolumn{3}{c}{$\textrm{Na}^{*}(n) + \textrm{Na}$}     & \multicolumn{3}{c}{$\textrm{H}^{*}(n) + \textrm{Li}$}    \\
   \cmidrule(lr){2-4} \cmidrule(lr){5-7} \cmidrule(lr){8-10} \cmidrule(lr){11-13}
n  & a1        & a2       & a3       & a1        & a2       & a3       & a1        & a2       & a3    & a1        & a2       & a3   \\ \hline \hline
4  & -35.12844 & 14.13288 & -1.89755 & -41.35531 & 18.67062 & -2.69225 & -46.51493 & 21.4718  & -3.10971 & -5.79278 & -1.1651  & 0.13442 \\
5  & -23.31012 & 7.66745  & -0.99756 & -16.97267 & 4.48624  & -0.58834 & -29.28982 & 11.44201 & -1.62003 & -6.22763 & -1.12647 & 0.12474 \\
6  & -17.64598 & 4.58023  & -0.56799 & -12.16141 & 1.71545  & -0.18066 & -21.22237 & 6.76472  & -0.92651 & -6.59811 & -1.12139 & 0.12283 \\
7  & -14.60487 & 2.93911  & -0.34127 & -12.25698 & 1.75651  & -0.18571 & -16.8308  & 4.22799  & -0.5506  & -6.96288 & -1.09353 & 0.11792 \\
8  & -12.93367 & 2.06542  & -0.22491 & -13.18087 & 2.32099  & -0.27765 & -14.3463  & 2.81506  & -0.34382 & -7.27818 & -1.07387 & 0.11462 \\
9  & -11.85235 & 1.5136   & -0.15519 & -13.53603 & 2.47414  & -0.30135 & -12.50968 & 1.74627  & -0.18362 & -7.54032 & -1.06697 & 0.11323 \\
10 & -11.29713 & 1.25562  & -0.12888 & -14.06446 & 2.72686  & -0.34263 & -11.58289 & 1.24937  & -0.11571 & -7.78576 & -1.05661 & 0.11177 \\
11 & -10.76294 & 0.97997  & -0.09759 & -14.21812 & 2.71937  & -0.34036 & -11.04281 & 0.97972  & -0.08309 & -7.96126 & -1.07659 & 0.11492 \\
12 & -10.38013 & 0.77342  & -0.07493 & -14.41123 & 2.74111  & -0.34371 & -10.72104 & 0.83367  & -0.06973 & -8.12687 & -1.09249 & 0.11762 \\
13 & -10.13548 & 0.63573  & -0.06144 & -14.78045 & 2.88417  & -0.36751 & -10.16092 & 0.51477  & -0.02736 & -8.35032 & -1.05952 & 0.11223 \\
14 & -9.85298  & 0.45989  & -0.0406  & -14.89837 & 2.86009  & -0.36292 & -10.36943 & 0.68833  & -0.06533 & -8.50237 & -1.06423 & 0.11285 \\
15 & -9.71227  & 0.36138  & -0.0305  & -15.367   & 3.08165  & -0.40038 & -9.68059  & 0.23985  & 0.00292  & -8.67836 & -1.04816 & 0.11043 \\
16 & -9.68919  & 0.32646  & -0.02892 & -15.22642 & 2.89562  & -0.36903 & -9.86564  & 0.37396  & -0.02657 & -8.75338 & -1.08765 & 0.11638 \\
17 & -9.56453  & 0.22389  & -0.01628 & -15.39519 & 2.92341  & -0.37371 & -9.64935  & 0.22401  & -0.00675 & -8.90127 & -1.07964 & 0.11542 \\
18 & -9.50279  & 0.15756  & -0.00879 & -15.54147 & 2.93973  & -0.3764  & -9.58519  & 0.17005  & -0.00234 & -9.01186 & -1.08791 & 0.11668 \\
19 & -9.48437  & 0.11619  & -0.00477 & -15.62221 & 2.91637  & -0.37249 & -9.51936  & 0.10875  & 0.00409  & -9.16473 & -1.06514 & 0.11307 \\
20 & -9.45545  & 0.06564  & 0.00121  & -15.80875 & 2.96859  & -0.38137 & -9.50371  & 0.07705  & 0.00612  & -9.23865 & -1.08876 & 0.11676
\end{tabular}
\end{table*}

\section{Results and Applications}
\label{sec:results}

\subsection{Rate coefficients for CI processes}

We calculated the partial $K^{(a)}(n,T)$ and $K^{(b)}(n,T)$
and total CI rate coefficients $K^{(ab)}(n,T)$
for the extended range of principal quantum numbers $n \le 25$ and temperatures
$500 \textrm{ K} \le T \le 1500$ K for the CI processes in $\textrm{Li}^{*}(n) + \textrm{Na}$,
$\textrm{Na}^{*}(n) + \textrm{Na}$ collisions, and $500 \textrm{ K} \le T \le 4000$ K for $\textrm{Li}^{*}(n) + \textrm{Li}$
and $\textrm{H}^{*}(n)+\textrm{Li}$ collisions. The total rate coefficients $K^{(ab)}(n,T)$ are presented in Figs.~\ref{Fig:2} and \ref{Fig:3} and in Tab.~\ref{tab:fit}. The rate coefficients $K^{(a)}(n,T)$ and $K^{(b)}(n,T)$ of both channels are presented in tabulated form Tabs.~\ref{Tab1} - \ref{Tab9}.

To enable better and more adequate usage of these results in laboratory as well as in modeling, we present a simple fitting formula for the total rate coefficients based on a least-square method, which represents a second-degree polynomial (logarithmic):
\begin{equation}
\log (K^{(ab)}(n;T))= a_{1}(n) + a_{2}(n)\cdot \log (T) + a_{3}(n)\cdot (\log (T))^{2}.
\label{eq:Fit}
\end{equation}
The fits are valid within the temperature range $500 \textrm{ K} \le T \le 1500$ K for $\textrm{Li}^{*}(n) + \textrm{Na}$,
$\textrm{Na}^{*}(n) + \textrm{Na}$ collisions and $500 \textrm{ K} \le T \le 4000$ K for $\textrm{Li}^{*}(n) + \textrm{Li}$ and $\textrm{H}^{*}(n) + \textrm{Li}$
collisions. It is possible that the fit is applicable outside this area, but this should be used with caution.
In Tab.~\ref{tab:fit} the selected fits (for $4 \le n \le 20$) for CI in A+RA collisions are listed.

The processes (a) and (b) in fact act as the channels of a more general CI process. Therefore we characterized them by partial rate coefficients
and coefficient $O^{(a)}(n,T)$, which describe the relative influence of the associative ionization channel. The calculations of those values are also performed.
The relative contribution of the associative channels (i.e. creation of molecular ions) are presented in
the bottom panels of Figs.~\ref{Fig:2} and \ref{Fig:3} by surface plot for the examples of $\textrm{Li}^{*}(n) + \textrm{Li}$ and $\textrm{H}^{*}(n) + \textrm{Li}$ collisions. It can be noticed that in the considered regions of $n$ and $T$ the process of associative CI in $\textrm{Li}^{*}(n) + \textrm{Li}$ collisions dominate in comparison with the non-associative CI channel with a maximum for small values of $n$. Similar conclusions are valid for the CI in $\textrm{Li}^{*}(n) + \textrm{Na}$, and $\textrm{Na}^{*}(n) + \textrm{Na}$ collisions (see Tabs. in online version of the paper).
Unlike these cases in $\textrm{H}^{*}(n) + \textrm{Li}$ collisions, the associative channel is
totally negligible (see Fig.~\ref{Fig:3}) for the analysed conditions e.g.
in Li-rich stars this process is definitely not the main source of the $\textrm{LiH}^{+}$ molecular ion creation.

\begin{figure*}
\begin{center}
\includegraphics[width=1.7\columnwidth]{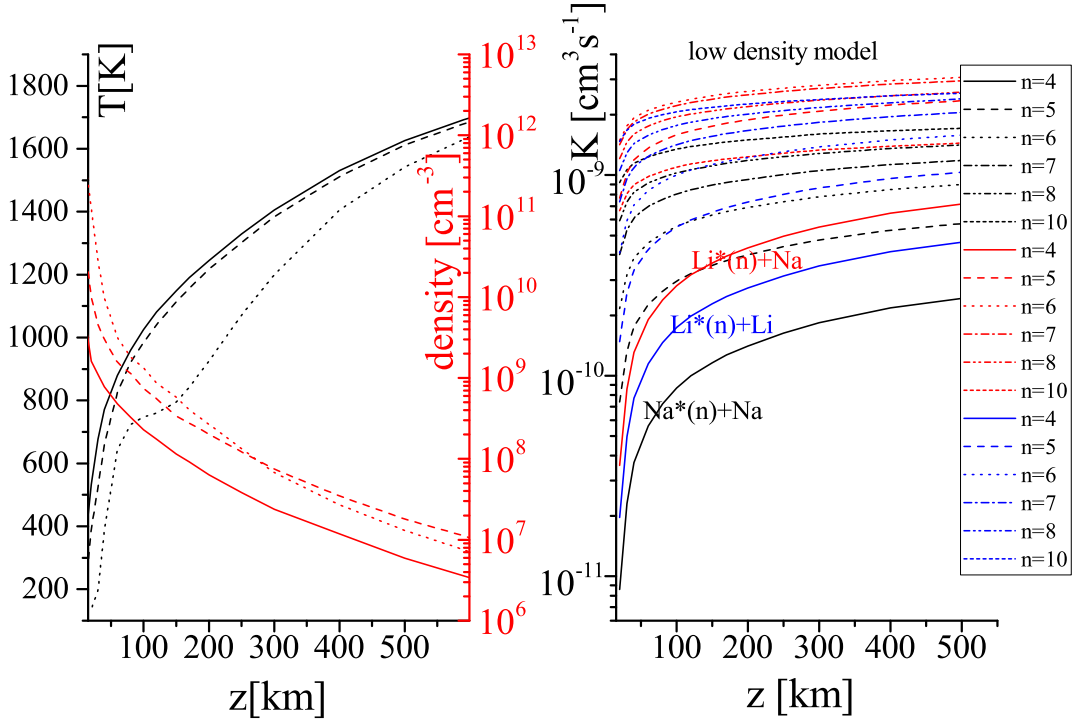}
\caption{Left: Temperature and density altitude profiles for low-density (solid line), moderate-density (dashed) and high-density model (dotted line) Io model atmosphere; Right: The total rate coefficient $K_{ci}^{(ab)}(n, T)$ for CI processes in $\textrm{Li}^{*}(n) + \textrm{Na}$, $ \textrm{Li}^{*}(n) + \textrm{Li}$ and $\textrm{Na}^{*}(n) + \textrm{Na}$ collisions for the density model atmosphere of Io \citep{str94}.}
\label{Fig:4}
\end{center}
\end{figure*}
Upper panels of Figs.~\ref{Fig:2} and \ref{Fig:3} present the total
CI rate coefficient $K^{(ab)}(n,T)$  for the cases of $\textrm{Li}^{*}(n) + \textrm{Li}$ and $\textrm{H}^{*}(n) + \textrm{Li}$
collisions.
One can see that the rate coefficient $K^{(ab)}(n,T)$ for $\textrm{Li}^{*}(n) + \textrm{Li}$ (and also for $\textrm{Li}^{*}(n) + \textrm{Na}$, and $\textrm{Na}^{*}(n) + \textrm{Na}$) has the maximum between $n=6$ and $n=10$ (upper panel of Fig.~\ref{Fig:2}).
When increasing the temperature, the maximum moves to lower values of $n$.
It is evidently that the total CI rate coefficient
for the higher values of $n$ does not or little depends on the temperature and can be
approximated with the function of $n$ i.e. $\sim 10^{-6}/n^{3}$ . A totally different behavior of $K^{(ab)}(n,T)$ is seen for the case of $\textrm{H}^{*}(n) + \textrm{Li}$
(see Fig.~\ref{Fig:3}) for which the rate coefficient monotony decreases with the increase of $n$, with different dependence on temperature.

\subsection{Io atmosphere}
The calculated CI rate coefficient $K^{(ab)}(n,T)$ (see Fig.~\ref{Fig:4}) covers the plasma parameter space which is
important for the models of Io's atmospheres \citep{str94}, which till now was not investigated in this context.
Such processes may become important and could be used for better numerical simulations and modelling.
In Fig~\ref{Fig:4} (left panel) the temperature and density altitude profiles for low density
(solid line), moderate-density (dashed line), and high-density (dotted line) of Io modeled atmospheres \citep{str94} are presented. The right panel of Fig~\ref{Fig:4} shows the total rate coefficient $K_{ci}^{(ab)}(n,T)$ for CI processes in $\textrm{Li}^{*}(n) + \textrm{Na}$, $\textrm{Li}^{*}(n) + \textrm{Li}$ and $\textrm{Na}^{*}(n) + \textrm{Na}$ collisions  for modeled atmospheres of Io \citep{str94,mos02}. One can notice the growth of coefficients at higher altitudes. Also, one can see that the rate coefficients increase with the increase of the principal quantum number n and that they are the largest for the Li*(n) + Na collisions. Additionally, they are larger for the Li*(n)+ Li than for the Na*(n) + Na collisions.
The investigation of  these processes is needed for better understanding of the interaction between the Io's atmosphere and Jovian magnetosphere \citep{wil02} and of the processes in the Jovian environment \citep{men90}. Although lithium has not been spectroscopically detected at Io to date, some authors \citep{sch05} expect its presence, and estimate the abundances of lithium as well as of other candidate elements (Rb, Cs, F, Br) at Io.

\subsection{Li-rich stars}
As it is known lithium is easily ionized (its ionization potential is 5.392 eV), thus Li I lines can be seen only in relatively cool stars with low effective temperatures i.e., stars from A to M spectral class. The Li I resonance line (670.78 nm) is most often observed in the spectra of cool stars. Most data on the lithium abundance in stars are obtained from the analysis of this line. The considered CI processes, which influence the ionization level and atom excited-state populations, could influence the optical properties and could be important for the lithium spectra.
Thus the CI collision processes could be of interest for lithium-rich stellar atmospheres \citep{sha01,sha03,nor98} as an additional channel for the creation of the neutral lithium atoms and atom excited-state populations. That is why we extended the range of temperatures (to include higher temperatures) for the calculation for lithium
(see Tabs. 4,5,8,9 and Figs. \ref{Fig:2} and \ref{Fig:3}) to enable the possible inclusion of CI processes in modeling of cool stars with low effective temperatures ($T_{\textrm{eff}} \le 6 000$ K) \citep[see e.g][]{kle16} and enormous high Li abundances e.g. subgiant J0741+2132 \citep[see][]{li18}.

Finally, the CI processes i.e. its associative channel (\ref{eq:nsim}(a) and \ref{eq:sim}(a)) in the binary collisions, leads to the formation of molecular ions,
giving a surprising variety of molecular formations in interstellar gas and can play an essential role \citep{dal76}. For understanding interstellar gas chemistry and for models of interstellar clouds the rate coefficients for reactions are needed as input parameter \citep{dis86}. The data for molecular ions LiNa$^{+}$, LiH$^{+}$ Li$_{2}^{+}$ and Na$_{2}^{+}$ are given in this contribution (see Tabs.~2,4,6,8).

The results of the rate coefficients of all investigated CI processes in this work are given in tabulated form in the online version of this article.

\section{Conclusions and Discussion}
The rate coefficients for the CI processes in $\textrm{Li}^{*}(n) + \textrm{Na}$,
$\textrm{Li}^{*}(n) + \textrm{Li}$, $\textrm{H}^{*}(n) + \textrm{Li}$ and $\textrm{Na}^{*}(n) + \textrm{Na}$
collisions were calculated. The obtained results have been applied to the models of atmosphere of Iovian satellite Io.
The presented values of the rate coefficients could be very useful for the improvement of modelling
and analysis of different layers of weakly ionized plasmas in atmospheres of various stars (lithium stars, photosphere of Sun, etc).
where these and other CI processes could be important and could change the optical characteristics.

We present the calculated rate coefficients of the corresponding
CI processes in the tabulated form, which is easy for further use, with a
particular accent to the applications for astro plasma research and low temperature laboratory plasma research
created in gas discharges, for example in microwave-induced discharges at atmospheric
pressure, where plasma conditions  may be favorable for processes investigated here.

In the near future we plan to further investigate the CI processes and develop methods
which could be applicable for the cases of extremely low temperatures which exists in
e.g. OGLE-2005-BLG-390Lb \citep{shv17} ultra cool Earth-mass planet orbiting dwarf.

\section*{Acknowledgements}
The authors are very grateful to Professor D. Ili\'c for support and fruitful discussion.
This work has been supported by the MESTD of the Republic of
Serbia Grants OI176002, III44002



\section*{SUPPORTING INFORMATION}
\label{Sec:suplement}

Supplementary data are available  online.
Additional Supporting Information (Tables 2-7) may be found in the
online version of this article. The tables are available in its entirety
for the principal quantum numbers $4 \le n\le 25$ and temperatures in machine-readable
form in the online journal as additional data.

\noindent Table 2. The rate coefficient $K^{(a)}(n,T)$[cm$^{3}$s$^{-1}$] for CI channel (\ref{eq:nsim1}), Li$^{*}$
(n)+Na.

\noindent Table 3. The rate coefficient $K^{(b)}(n,T)$[cm$^{3}$s$^{-1}$] for CI channel (\ref{eq:nsim2}), Li$^{*}$
(n)+Na.

\noindent Table 4. The rate coefficient $K^{(a)}(n,T)$[cm$^{3}$s$^{-1}$] for CI channel (\ref{eq:sim1}), Li$^{*}$
(n)+Li.

\noindent Table 5. The rate coefficient $K^{(b)}(n,T)$[cm$^{3}$s$^{-1}$] for CI channel (\ref{eq:sim2}), Li$^{*}$
(n)+Li.

\noindent Table 6. The rate coefficient $K^{(a)}(n,T)$[cm$^{3}$s$^{-1}$] for CI channel (\ref{eq:sim1}), Na$^{*}$
(n)+Na.

\noindent Table 7. The rate coefficient $K^{(b)}(n,T)$[cm$^{3}$s$^{-1}$] for CI channel (\ref{eq:sim2}), Na$^{*}$
(n)+Na.

\noindent Table 8. The rate coefficient $K^{(a)}(n,T)$[cm$^{3}$s$^{-1}$] for CI channel (\ref{eq:nsim1}), H$^{*}$
(n)+Li.

\noindent Table 9. The rate coefficient $K^{(b)}(n,T)$[cm$^{3}$s$^{-1}$] for CI channel (\ref{eq:nsim2}), H$^{*}$
(n)+Li.

\begin{table*}
\caption{The rate coefficient $K^{(a)}(n,T)$[cm$^{3}$s$^{-1}$] for CI channel (\ref{eq:nsim1}), Li$^{*}$
(n)+Na. This table is available in its entirety in machine-readable form in the online journal as additional data.
A portion is shown here for guidance regarding its form and content.}
\label{Tab1}
\begin{tabular}{llllllllllll}
\hline
    \multicolumn{11}{c}{T[K]}&\\
    \cline{2-12}
$n$ & 500      & 600      & 700      & 800      & 900      & 1000     & 1100     & 1200     & 1300     & 1400     & 1500     \\
\hline \hline
4   & 2.61E-11 & 5.52E-11 & 9.53E-11 & 1.44E-10 & 1.98E-10 & 2.56E-10 & 3.16E-10 & 3.75E-10 & 4.33E-10 & 4.89E-10 & 5.40E-10 \\
10  & 6.24E-10 & 7.19E-10 & 7.98E-10 & 8.63E-10 & 9.15E-10 & 9.57E-10 & 9.89E-10 & 1.01E-09 & 1.03E-09 & 1.05E-09 & 1.06E-09 \\
15  & 1.07E-10 & 1.25E-10 & 1.39E-10 & 1.51E-10 & 1.60E-10 & 1.68E-10 & 1.74E-10 & 1.79E-10 & 1.82E-10 & 1.85E-10 & 1.87E-10 \\
20  & 2.63E-11 & 3.06E-11 & 3.41E-11 & 3.70E-11 & 3.94E-11 & 4.13E-11 & 4.28E-11 & 4.39E-11 & 4.48E-11 & 4.55E-11 & 4.59E-11 \\
25  & 8.69E-12 & 1.01E-11 & 1.13E-11 & 1.22E-11 & 1.30E-11 & 1.36E-11 & 1.41E-11 & 1.45E-11 & 1.48E-11 & 1.50E-11 & 1.52E-11
\end{tabular}
\end{table*}
\begin{table*}
\caption{The rate coefficient $K^{(b)}(n,T)$[cm$^{3}$s$^{-1}$] for CI channel (\ref{eq:nsim2}), Li$^{*}$
(n)+Na. This table is available in its entirety in machine-readable form in the online journal as additional data.
A portion is shown here for guidance regarding its form and content.}
\label{Tab2}
\begin{tabular}{llllllllllll}
\hline
    \multicolumn{11}{c}{T[K]}&\\
    \cline{2-12}
$n$ & 500      & 600      & 700      & 800      & 900      & 1000     & 1100     & 1200     & 1300     & 1400     & 1500     \\
\hline \hline
4   & 0.00E+00 & 0.00E+00 & 4.45E-15 & 9.04E-13 & 3.29E-12 & 7.80E-12 & 1.51E-11 & 2.59E-11 & 4.05E-11 & 5.87E-11 & 8.18E-11 \\
10  & 6.79E-12 & 1.75E-11 & 3.44E-11 & 5.75E-11 & 8.61E-11 & 1.19E-10 & 1.56E-10 & 1.96E-10 & 2.38E-10 & 2.82E-10 & 3.27E-10 \\
15  & 1.13E-12 & 2.91E-12 & 5.74E-12 & 9.60E-12 & 1.44E-11 & 2.00E-11 & 2.62E-11 & 3.29E-11 & 3.99E-11 & 4.73E-11 & 5.48E-11 \\
20  & 2.75E-13 & 7.12E-13 & 1.40E-12 & 2.35E-12 & 3.52E-12 & 4.89E-12 & 6.41E-12 & 8.04E-12 & 9.77E-12 & 1.16E-11 & 1.34E-11 \\
25  & 9.08E-14 & 2.35E-13 & 4.63E-13 & 7.75E-13 & 1.16E-12 & 1.61E-12 & 2.11E-12 & 2.65E-12 & 3.22E-12 & 3.82E-12 & 4.43E-12
\end{tabular}
\end{table*}

\begin{table*}
\caption{The rate coefficient $K^{(a)}(n,T)$[cm$^{3}$s$^{-1}$] for CI channel (\ref{eq:sim1}), Li$^{*}$
(n)+Li. This table is available in its entirety in machine-readable form in the online journal as additional data.
A portion is shown here for guidance regarding its form and content.}
\label{Tab3}
\begin{tabular}{llllllllllll}
\hline
    \multicolumn{11}{c}{T[K]}&\\
    \cline{2-12}
$n$ & 500      & 700      & 900      & 1000      & 1300      & 1500     & 2000     & 2500     & 3000     & 3500     & 4000     \\
\hline \hline
4  & 1.40E-11 & 5.53E-11 & 1.20E-10 & 1.58E-10 & 2.74E-10 & 3.47E-10 & 4.84E-10 & 5.63E-10 & 6.00E-10 & 6.00E-10 & 6.00E-10 \\
10 & 1.20E-09 & 1.25E-09 & 1.23E-09 & 1.20E-09 & 1.13E-09 & 1.08E-09 & 9.55E-10 & 8.54E-10 & 7.69E-10 & 6.99E-10 & 6.41E-10 \\
15 & 7.46E-10 & 6.53E-10 & 5.78E-10 & 5.47E-10 & 4.70E-10 & 4.30E-10 & 3.53E-10 & 3.00E-10 & 2.60E-10 & 2.29E-10 & 2.04E-10 \\
20 & 2.96E-10 & 2.42E-10 & 2.05E-10 & 1.91E-10 & 1.58E-10 & 1.42E-10 & 1.13E-10 & 9.42E-11 & 8.02E-11 & 6.99E-11 & 6.17E-11 \\
25 & 1.21E-10 & 9.54E-11 & 7.93E-11 & 7.33E-11 & 5.97E-11 & 5.32E-11 & 4.17E-11 & 3.43E-11 & 2.91E-11 & 2.52E-11 & 2.21E-11
\end{tabular}
\end{table*}
\begin{table*}
\caption{The rate coefficient $K^{(b)}(n,T)$[cm$^{3}$s$^{-1}$] for CI channel (\ref{eq:sim2}), Li$^{*}$
(n)+Li. This table is available in its entirety in machine-readable form in the online journal as additional data.
A portion is shown here for guidance regarding its form and content.}
\label{Tab4}
\begin{tabular}{llllllllllll}
\hline
    \multicolumn{11}{c}{T[K]}&\\
    \cline{2-12}
$n$ & 500      & 700      & 900      & 1000      & 1300      & 1500     & 2000     & 2500     & 3000     & 3500     & 4000     \\
\hline \hline
4  & 0.00E+00 & 0.00E+00 & 1.99E-12 & 4.86E-12 & 2.61E-11 & 5.31E-11 & 1.70E-10 & 3.39E-10 & 5.36E-10 & 7.52E-10 & 9.62E-10 \\
10 & 2.18E-10 & 4.61E-10 & 7.14E-10 & 8.37E-10 & 1.17E-09 & 1.38E-09 & 1.82E-09 & 2.17E-09 & 2.47E-09 & 2.72E-09 & 2.94E-09 \\
15 & 3.52E-10 & 5.16E-10 & 6.47E-10 & 7.03E-10 & 8.41E-10 & 9.15E-10 & 1.06E-09 & 1.16E-09 & 1.24E-09 & 1.30E-09 & 1.36E-09 \\
20 & 2.43E-10 & 3.08E-10 & 3.55E-10 & 3.74E-10 & 4.18E-10 & 4.41E-10 & 4.83E-10 & 5.11E-10 & 5.33E-10 & 5.50E-10 & 5.63E-10 \\
25 & 1.44E-10 & 1.70E-10 & 1.88E-10 & 1.95E-10 & 2.10E-10 & 2.18E-10 & 2.33E-10 & 2.42E-10 & 2.49E-10 & 2.55E-10 & 2.59E-10
\end{tabular}
\end{table*}

\begin{table*}
\caption{The rate coefficient $K^{(a)}(n,T)$[cm$^{3}$s$^{-1}$] for CI channel (\ref{eq:sim1}), Na$^{*}$
(n)+Na. This table is available in its entirety in machine-readable form in the online journal as additional data.
A portion is shown here for guidance regarding its form and content.}
\label{Tab5}
\begin{tabular}{llllllllllll}
\hline
    \multicolumn{11}{c}{T[K]}&\\
    \cline{2-12}
$n$ & 500      & 600      & 700      & 800      & 900      & 1000     & 1100     & 1200     & 1300     & 1400     & 1500     \\
\hline \hline
4   & 5.96E-12 & 1.39E-11 & 2.58E-11 & 4.09E-11 & 5.85E-11 & 7.78E-11 & 9.81E-11 & 1.19E-10 & 1.39E-10 & 1.58E-10 & 1.77E-10 \\
10  & 7.05E-10 & 7.19E-10 & 7.19E-10 & 7.11E-10 & 6.96E-10 & 6.81E-10 & 6.64E-10 & 6.46E-10 & 6.29E-10 & 6.14E-10 & 5.96E-10 \\
15  & 6.32E-10 & 5.95E-10 & 5.60E-10 & 5.29E-10 & 5.00E-10 & 4.75E-10 & 4.52E-10 & 4.31E-10 & 4.12E-10 & 3.94E-10 & 3.79E-10 \\
20  & 2.98E-10 & 2.69E-10 & 2.46E-10 & 2.26E-10 & 2.10E-10 & 1.96E-10 & 1.83E-10 & 1.73E-10 & 1.63E-10 & 1.54E-10 & 1.47E-10 \\
25  & 1.30E-10 & 1.15E-10 & 1.03E-10 & 9.41E-11 & 8.63E-11 & 7.97E-11 & 7.41E-11 & 6.93E-11 & 6.51E-11 & 6.13E-11 & 5.80E-11
\end{tabular}
\end{table*}
\begin{table*}
\caption{The rate coefficient $K^{(b)}(n,T)$[cm$^{3}$s$^{-1}$] for CI channel (\ref{eq:sim2}), Na$^{*}$
(n)+Na. This table is available in its entirety in machine-readable form in the online journal as additional data.
A portion is shown here for guidance regarding its form and content.}
\label{Tab6}
\begin{tabular}{llllllllllll}
\hline
    \multicolumn{11}{c}{T[K]}&\\
    \cline{2-12}
$n$ & 500      & 600      & 700      & 800      & 900      & 1000     & 1100     & 1200     & 1300     & 1400     & 1500     \\
\hline \hline
4   & 0.00E+00 & 0.00E+00 & 0.00E+00 & 3.53E-13 & 1.31E-12 & 3.09E-12 & 6.01E-12 & 1.02E-11 & 1.60E-11 & 2.36E-11 & 3.28E-11 \\
10  & 1.77E-10 & 2.69E-10 & 3.66E-10 & 4.63E-10 & 5.59E-10 & 6.51E-10 & 7.39E-10 & 8.24E-10 & 9.05E-10 & 9.82E-10 & 1.06E-09  \\
15  & 3.39E-10 & 4.24E-10 & 5.01E-10 & 5.70E-10 & 6.33E-10 & 6.89E-10 & 7.40E-10 & 7.88E-10 & 8.31E-10 & 8.72E-10 & 9.08E-10 \\
20  & 2.63E-10 & 3.03E-10 & 3.36E-10 & 3.64E-10 & 3.89E-10 & 4.10E-10 & 4.29E-10 & 4.46E-10 & 4.61E-10 & 4.75E-10 & 4.88E-10 \\
25  & 1.63E-10 & 1.80E-10 & 1.93E-10 & 2.04E-10 & 2.14E-10 & 2.22E-10 & 2.29E-10 & 2.35E-10 & 2.41E-10 & 2.46E-10 & 2.51E-10
\end{tabular}
\end{table*}

\begin{table*}
\caption{The rate coefficient $K^{(a)}(n,T)$[cm$^{3}$s$^{-1}$] for CI channel (\ref{eq:nsim1}), H$^{*}$
(n)+Li. This table is available in its entirety in machine-readable form in the online journal as additional data.
A portion is shown here for guidance regarding its form and content.}
\label{Tab8}
\begin{tabular}{llllllllllll}
\hline
    \multicolumn{11}{c}{T[K]}&\\
    \cline{2-12}
$n$ & 500      & 700      & 900      & 1000      & 1300      & 1500     & 2000     & 2500     & 3000     & 3500     & 4000     \\
\hline \hline
4  & 8.80E-11 & 7.75E-11 & 6.70E-11 & 6.18E-11 & 5.47E-11 & 4.99E-11 & 4.28E-11 & 3.79E-11 & 3.41E-11 & 3.11E-11 & 2.86E-11 \\
10 & 1.21E-12 & 1.07E-12 & 9.22E-13 & 8.50E-13 & 7.50E-13 & 6.84E-13 & 5.85E-13 & 5.17E-13 & 4.65E-13 & 4.22E-13 & 3.88E-13 \\
15 & 1.60E-13 & 1.41E-13 & 1.22E-13 & 1.12E-13 & 9.90E-14 & 9.03E-14 & 7.73E-14 & 6.83E-14 & 6.13E-14 & 5.57E-14 & 5.12E-14 \\
20 & 3.81E-14 & 3.35E-14 & 2.89E-14 & 2.66E-14 & 2.35E-14 & 2.14E-14 & 1.83E-14 & 1.62E-14 & 1.46E-14 & 1.32E-14 & 1.22E-14 \\
25 & 1.25E-14 & 1.10E-14 & 9.48E-15 & 8.73E-15 & 7.71E-15 & 7.03E-15 & 6.01E-15 & 5.31E-15 & 4.77E-15 & 4.33E-15 & 3.98E-15
\end{tabular}
\end{table*}
\begin{table*}
\caption{The rate coefficient $K^{(b)}(n,T)$[cm$^{3}$s$^{-1}$] for CI channel (\ref{eq:nsim2}), H$^{*}$
(n)+Li. This table is available in its entirety in machine-readable form in the online journal as additional data.
A portion is shown here for guidance regarding its form and content.}
\label{Tab9}
\begin{tabular}{llllllllllll}
\hline
    \multicolumn{11}{c}{T[K]}&\\
    \cline{2-12}
$n$ & 500      & 700      & 900      & 1000      & 1300      & 1500     & 2000     & 2500     & 3000     & 3500     & 4000     \\
\hline \hline
4  & 1.09E-08 & 9.87E-09 & 8.84E-09 & 8.33E-09 & 7.65E-09 & 7.20E-09 & 6.65E-09 & 6.30E-09 & 6.01E-09 & 5.78E-09 & 5.67E-09 \\
10 & 1.49E-10 & 1.34E-10 & 1.19E-10 & 1.12E-10 & 1.02E-10 & 9.53E-11 & 8.73E-11 & 8.19E-11 & 7.76E-11 & 7.38E-11 & 7.21E-11 \\
15 & 1.97E-11 & 1.77E-11 & 1.58E-11 & 1.48E-11 & 1.35E-11 & 1.26E-11 & 1.15E-11 & 1.08E-11 & 1.02E-11 & 9.74E-12 & 9.51E-12 \\
20 & 4.67E-12 & 4.20E-12 & 3.73E-12 & 3.50E-12 & 3.19E-12 & 2.99E-12 & 2.73E-12 & 2.57E-12 & 2.43E-12 & 2.31E-12 & 2.26E-12 \\
25 & 1.53E-12 & 1.38E-12 & 1.23E-12 & 1.15E-12 & 1.05E-12 & 9.78E-13 & 8.96E-13 & 8.41E-13 & 7.97E-13 & 7.57E-13 & 7.40E-13
\end{tabular}
\end{table*}

\end{document}